%% file: main.tex
\documentclass[sigconf]{acmart}

\AtBeginDocument{%
  \providecommand\BibTeX{{%
    \normalfont B\kern-0.5em{\scshape i\kern-0.25em b}\kern-0.8em\TeX}}}

\copyrightyear{2024}
\acmYear{2024}
\setcopyright{acmlicensed}
\acmConference[WWW '24]{Proceedings of the ACM
Web Conference 2024}{May 13--17, 2024}{Singapore, Singapore}
\acmBooktitle{Proceedings of the ACM Web Conference 2024 (WWW '24), May
13--17, 2024, Singapore, Singapore}
\acmDOI{10.1145/3589334.3645622}
\acmISBN{979-8-4007-0171-9/24/05}
\usepackage{enumitem}
\usepackage{booktabs,siunitx}
\usepackage{subfigure}
\newcommand{\mc}[3]{\multicolumn{#1}{#2}{#3}}
\newcommand{\vpara}[1]{\vspace{0.05in}\noindent \textbf{#1 }}

\usepackage[ruled,vlined]{algorithm2e}

\SetCommentSty{mycommfont}

\SetKwInput{KwInput}{Input}                
\SetKwInput{KwOutput}{Output}              

\acmSubmissionID{1847}

\begin{document}

\title[On the Feasibility of Simple Transformer for Dynamic Graph Modeling]{On the Feasibility of Simple Transformer \\ for Dynamic Graph Modeling}

\author{Yuxia Wu}
\affiliation{%
    \institution{Singapore Management University}
    \country{Singapore} 
}
\email{yuxiawu@smu.edu.sg}

\author{Yuan Fang*}
\affiliation{%
    \institution{Singapore Management University}
    \country{Singapore} 
}
\email{yfang@smu.edu.sg}

\author{Lizi Liao}
\affiliation{%
    \institution{Singapore Management University}
    \country{Singapore} 
}
\email{lzliao@smu.edu.sg}

\thanks{
    * Corresponding author.
}

\renewcommand{\shortauthors}{Yuxia Wu, Yuan Fang, and Lizi Liao}

\begin{abstract}
Dynamic graph modeling is crucial for understanding complex structures in web graphs, spanning applications in social networks, recommender systems, and more. Most existing methods primarily emphasize structural dependencies and their temporal changes. However, these approaches often overlook detailed temporal aspects or struggle with long-term dependencies. Furthermore, many solutions overly complicate the process by emphasizing intricate module designs to capture dynamic evolutions.
In this work, we harness the strength of the Transformer's self-attention mechanism, known for adeptly handling long-range dependencies in sequence modeling. Our approach offers a simple Transformer model, called SimpleDyG, tailored for dynamic graph modeling without complex modifications. We re-conceptualize dynamic graphs as a sequence modeling challenge and introduce a novel temporal alignment technique. This technique not only captures the inherent temporal evolution patterns within dynamic graphs but also streamlines the modeling process of their evolution. To evaluate the efficacy of SimpleDyG, we conduct extensive experiments on four real-world datasets from various domains. The results demonstrate the competitive performance of SimpleDyG in comparison to a series of state-of-the-art approaches despite its simple design.  

\end{abstract}


\maketitle

\input{intro}

\input{related}

\input{pre}

\input{method}
\input{exp}

\section{Conclusion}
In this work, we explored the problem of dynamic graph modeling, recognizing its significance across a wide range of applications. Drawing from the strengths of the Transformer's self-attention mechanism, we tailored a solution that supersedes the often convoluted designs in many existing methods. Our novel approach, named SimpleDyG, reformulates dynamic graphs from a sequence modeling perspective. It is superior to not only the discrete-time approaches by considering the full temporal order, but also the continuous-time approaches by capturing long-term dependencies using a Transformer. SimpleDyG is exceptionally simple as it does not modify the Transformer architecture; instead, it maps a dynamic graph into a set of sequences via temporal ego-graphs, and modifies the input sequences to achieve temporal alignment. Nevertheless, SimpleDyG achieves surprisingly strong performance in diverse dynamic graphs.
As future work, we will investigate the nuances of the temporal alignment technique for further optimizations.

\section*{Acknowledgments}
This research / project is supported by the Ministry of Education, Singapore, under its Academic Research Fund Tier 2 (Proposal ID: T2EP20122-0041). Any opinions, findings and conclusions or recommendations expressed in this material are those of the author(s) and
do not reflect the views of the Ministry of Education, Singapore.

\bibliographystyle{ACM-Reference-Format}
\bibliography{sample-base}

\appendix
\numberwithin{equation}{section}
\numberwithin{figure}{section}
\newpage

\input{app}

\end{document}

%% file: intro.tex
\section{Introduction}

\begin{figure*}[ht] 
\centering
\subfigure[Discrete-time approaches]{
\centering
    \includegraphics[scale=0.4]{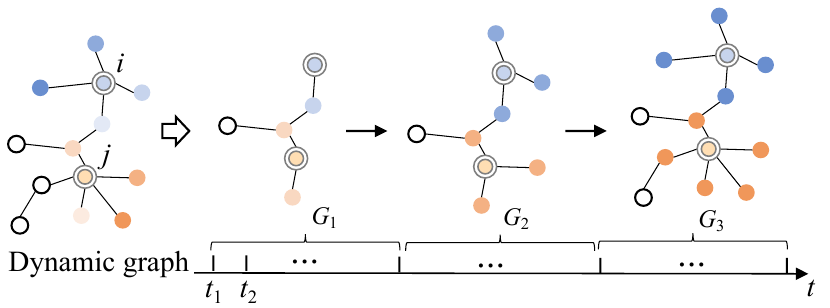}
    \label{discrete}
}
\subfigure[Continuous-time approaches]{
    \centering
    \includegraphics[scale=0.4]{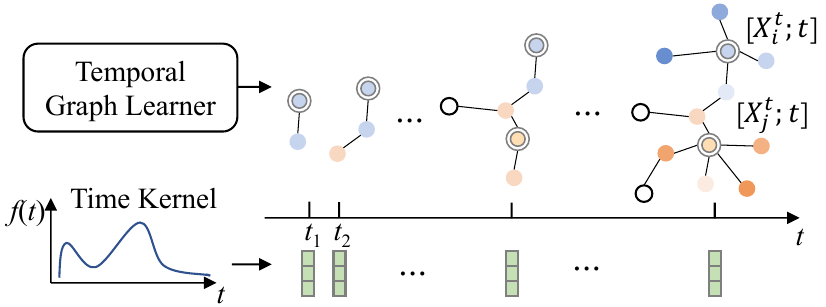}
    \label{continuous}

}
~~~
\subfigure[Self-attention in Transformer]{
    \centering
    \includegraphics[scale=0.4]{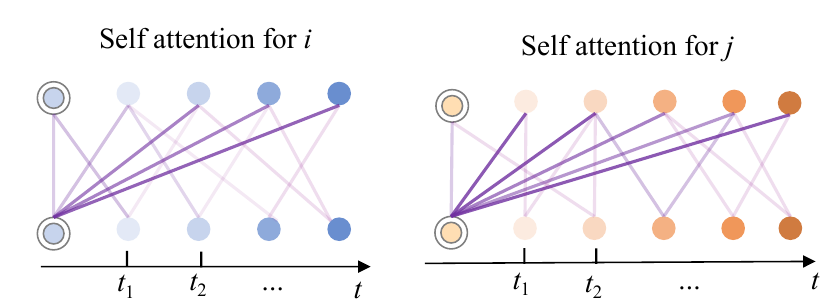}
    \label{self-attention}

}
\vspace{-0.4cm}
 \caption{Dynamic graph modeling can be summarized as follows: (a) Discrete-time methods treat the dynamic graph as a series of snapshots, ignoring the timing details within each. (b) Continuous-time methods factor in the timing of interactions, using them along with a graph learning process to update node representations $X_i^t$ at each time $t$. (c) Transformer-based models handle node sequences continuously, utilizing self-attention to recognize long-term dependencies.}
\end{figure*}

Graph-structured data are prevalent on the World Wide Web \cite{yu2024few}, such as social networks \cite{perozzi2014deepwalk,fan2019graph}, recommender systems \cite{wu2020personalized,wu2022state}, citation graphs \cite{kanakia2019scalable,yu2023hgprompt,yu2023multigprompt}, dialogue systems \cite{wu2022semi,liao2023proactive}, and so on. Thus, graph-based mining and learning have become fundamental tools in many Web applications, ranging from analyzing communication patterns within social friendships, to predicting users' behavior on digital platforms and investigating citation trends in the academic community.  
Traditionally, many works focus on static graphs characterized by fixed nodes and edges. However, many real-world graphs on the Web are intrinsically dynamic, which continuously evolve over time \cite{sankar2020dysat}. 
That is, the nodes and their edges in such graphs are undergoing constant addition or reorganization based on some underlying patterns of evolution. For example, in a social network like UCI \cite{panzarasa2009patterns}, where nodes represent users and edges represent messages exchanged between users, new edges constantly emerge as 
users frequently exchange messages with their friends.
To study this important class of graphs and their applications on the Web, we focus on dynamic graph modeling in this paper, aiming to capture the evolving patterns in a dynamic graph.

Existing works for dynamic graph modeling mainly fall into two categories: discrete-time approaches \cite{sankar2020dysat,pareja2020evolvegcn} and continuous-time approaches \cite{trivedi2019dyrep,xu2020inductive,wen2022trend,cong2022we},  as shown in Figures \ref{discrete} and \ref{continuous}, respectively. The former regards dynamic graphs as a sequence of snapshots over a discrete set of time steps. This kind of approach usually leverages structural modules such as graph neural networks (GNN) \cite{wu2020comprehensive} to capture the topological information of graphs, and temporal modules such as recurrent neural networks (RNN) \cite{schuster1997bidirectional} to learn the sequential evolution of the snapshots \cite{sankar2020dysat}. Meanwhile, the latter focuses on modeling continuous temporal patterns via specific temporal modules such as temporal random walk \cite{nguyen2018continuous} or temporal kernel \cite{de1992gamma}, illustrated by Figure \ref{continuous}. Despite the significant progress made in dynamic graph modeling, there still exist some key limitations. First, the modeling of temporal dynamics on graphs is still coarse-grained or short-termed. On one hand, discrete-time approaches discard the fine-grained temporal information within the snapshot, which inevitably results in partial loss of temporal patterns. On the other hand, while continuous-time approaches retain full temporal details by mapping each interaction to a continuous temporal space, capturing long-term dependency within historical graph data still remains a difficult problem \cite{rossi2020temporal,yu2023towards}. Second, the majority of the existing works rely extensively on the message-passing GNNs to encode the structural patterns in dynamic graphs. Although powerful in graph modeling, the message-passing mechanism shows inherent limitations such as over-smoothing \cite{chen2020measuring} and over-squashing \cite{alon2020bottleneck} that become more pronounced as model depth increases, preventing deeper and more expressive architectures.

In pursuit of addressing these limitations, we have been intrigued by the successful application of Transformer \cite{vaswani2017attention} and its variants in natural language processing (NLP) \cite{kenton2019bert,brown2020language,liao2021dialogue} and computer vision (CV) \cite{dosovitskiy2020image,liu2021swin}. The success is underpinned by two distinct advantages inherent to the Transformer architecture: as shown in Figure \ref{self-attention}, it can naturally support a continuous sequence of data without the need for discrete snapshots, and its self-attention mechanism can capture long-term dependency \cite{vaswani2017attention}, which are important factors for dynamic graph modeling. Transformer also presents a potentially better alternative to capturing topological information, as it is less affected by the over-smoothing and over-squashing issues associated with message-passing GNNs. Hence, in this work, we explore the feasibility of the Transformer architecture for dynamic graph modeling. In fact, we have observed a growing body of research trying to modify the Transformer for static graphs  \cite{ying2021transformers, rampavsek2022recipe, kim2022pure}. Nonetheless, these studies primarily focus on integrating graph structural knowledge into the vanilla Transformer model, which generally still leverage message-passing GNNs as auxiliary modules to refine positional encoding and attention matrices based on graph-derived information \cite{min2022transformer}. More recently, \citet{ying2021transformers} have demonstrated that the pure Transformer architecture holds promise for graphs. However, all these previous Transformer-based approaches only focus on static graphs, leaving unanswered questions about the feasibility for dynamic graphs, as follows. 

The first challenge lies in the need to preserve the historical evolution throughout the entire timeline.
However, due to the calculation of pairwise attention scores, existing Transformer-based graph models can only deal with a small receptive field, and would face serious scalability issues on even a moderately large graph.
Notably, their primary application is limited to small-size graphs such as molecular graphs \cite{rampavsek2022recipe}. 
However, many dynamic graphs on the Web such as social networks or citation graphs are generally much larger for the vanilla Transformer to handle. 
To this end, we adopt a novel strategy wherein we treat the history of each node as a \emph{temporal ego-graph}, serving as the receptive field of the ego-node. The temporal ego-graph is much smaller than the entire graph, yet it retains the full interaction history of the ego-node in the dynamic graph. Thus, we are able to preserve the temporal dynamics of every user across the entire timeline, while simultaneously ensuring scalability.
Subsequently, this temporal ego-graph can be tokenized into a sequential input tailored for the Transformer. Remarkably, through this simple tokenization process, no modification to the original Transformer architecture is required.

The second challenge lies in the need to align temporal information across input sequences. Specifically, on dynamic graphs different input sequences converge within a common time domain---whether through absolute points in time (e.g., 10am on 12 October 2023) or relative time intervals (e.g., a one-hour window), with uniformity across all sequences generated from different nodes' history. In contrast, sequences for natural language or static graphs lack such a universal time domain, and can be regarded as largely independent of each other. Thus, vanilla sequences without temporal alignment lack a way to differentiate variable time intervals and frequency information. For example, a bursty stream of interactions, happening over a short one-hour interval, has a distinct evolution pattern from a steady stream containing the same number of interactions, but happening over a period of one day.

Therefore, it becomes imperative to introduce a mechanism that infuses temporal alignment among distinct input sequences generated from the ego-graphs. To address this challenge, we carefully design special \emph{temporal tokens} to align different input sequences in the time domain. The temporal tokens serve as indicators of distinct time steps that are globally recognized across all nodes,
thereby unifying different input sequences. While achieving the global alignment, local sequences from each node still retain the chronological order of the interactions in-between the temporal tokens, unlike traditional discrete-time approaches where temporal information within each snapshot is lost.  


Based on the above insights, we propose a \textbf{Simple} Transformer architecture for \textbf{Dy}namic \textbf{G}raph modeling, named \textbf{SimpleDyG}. The word ``simple'' is a reference to the use of the original Transformer architecture without any modification, where the capability of dynamic graph modeling is simply and solely derived from constructing and modifying the input sequences. 
In summary, the contribution of our work is threefold. %
(1) We explore the potential of the Transformer architecture for modeling dynamic graphs. We propose a simple yet surprisingly effective Transformer-based approach for dynamic graphs, called SimpleDyG, without complex modifications. 
(2) We introduce a novel strategy to map a dynamic graph into a set of sequences to improve the scalability, by considering the history of each node as a temporal ego-graph. Furthermore, we design special temporal tokens to achieve global temporal alignment across nodes, yet preserving the chronological order of interactions at a local level.
(3) We conduct extensive experiments and analyses across four real-world Web graphs, spanning diverse domains on the Web. The empirical results demonstrate not only the feasibility, but also the superiority of SimpleDyG.

%% file: related.tex
\section{Related Work}

\vpara{Dynamic Graph Learning.}
Current dynamic graph learning methods can be categorized into discrete-time and continuous-time approaches. In discrete-time methods, dynamic graphs are treated as a series of static graph snapshots taken at regular time intervals. To model both structural and temporal aspects, these approaches integrate the GNNs with sequence models, such as RNNs or self-attention mechanisms \cite{sankar2020dysat,pareja2020evolvegcn,goyal2020dyngraph2vec, taheri2019learning}. For instance, DySAT \cite{sankar2020dysat} leverages a graph attention network and self-attention as fundamental components for both structural and temporal modules, whereas EvolveGCN \cite{pareja2020evolvegcn} employs an RNN to evolve the parameters of graph convolutional networks. Nevertheless, they often fall short of capturing the granular temporal information. In contrast, the continuous-time approaches consider every interaction event at each specific timestamp. Some approaches model dynamic graph evolution as temporal random walks or causal anonymous walks \cite{nguyen2018continuous,wang2021inductive}. Another line of research focuses on time encoding techniques, which integrate with graph structure modeling, such as the temporal graph attention used in TGAT \cite{xu2020inductive} and TGN \cite{rossi2020temporal}, or MLP-Mixer layers applied in GraphMixer \cite{cong2022we}. Additionally, researchers also leverage temporal point processes to capture the graph formation process \cite{trivedi2019dyrep, ji2021dynamic, wen2022trend}. Despite the promise demonstrated by continuous-time approaches, it is important to note that they come with limitations in capturing long-term dependencies within historical data.

The differences between our work and the previous dynamic graph learning methods lie in two points. First, our method effectively mitigates the long-term dependency challenge, leveraging the inherent advantages of the Transformer architecture. Second, our method preserves the chronological history of each input sequence. In particular, the temporal alignment mechanism synchronizes different input sequences, empowering our model to capture both global and local information within the dynamic graphs. 


\vpara{Transformers for Graphs.} 
Transformer architectures for graphs have emerged as a compelling alternative to conventional GNNs, aiming to mitigate issues like over-smoothing and over-squashing. Prior research has focused on integrating graph information into the vanilla Transformer through diverse strategies. Some methods integrate GNNs as auxiliary components to bolster structural comprehension within the Transformer architecture \cite{rong2020self, kim2021transformers}. 
Others focus on enriching positional embeddings by spatial information derived from the graph. For 
instance, Graphormer \cite{ying2021transformers} integrates the centrality, spatial and edge encoding into Transformers, whereas \citet{cai2020graph} adopt distance embedding for tree-structured abstract meaning representation graph, and \citet{kreuzer2021rethinking} utilize the full Laplacian spectrum to learn the positional encoding.   
There are also studies on refining the attention mechanism in Transformers for graph modeling. For instance, \citet{min2022neighbour} employ a graph masking attention mechanism to seamlessly inject graph-related priors into the Transformer architecture. More recently, \citet{kim2022pure} have shed light on the effectiveness of pure Transformers in graph learning without complex designs. Their approach treats all nodes and edges as independent tokens, serving as inputs for the Transformer. 
Besides, \citet{mao2023hinormer} propose a Transformer-based model for heterogeneous graphs, integrating local structures and heterogeneous relations into the attention mechanism. 

It is worth noting that most of the previous works based on Transformers mainly deal with static graphs. Recently, \citet{yu2023towards} have introduced a Transformer-based model designed for dynamic graph learning, which is contemporary with our work. The difference lies in that they rely on complex designs for capturing co-occurrence neighbors of different nodes and encoding temporal intervals. In contrast, we explore a simple Transformer for dynamic graphs without the need for complex modifications.

%% file: pre.tex
\section{PRELIMINARIES}
We first define the problem of dynamic graph modeling. Then, we briefly introduce the background of the Transformer architecture.

\subsection{Dynamic Graph Modeling}

We define a dynamic graph as $\mathcal{G}=(\mathcal{V},\mathcal{E}, \mathcal{T}, \mathcal{X})$ with a set of nodes $\mathcal{V}$, edges $\mathcal{E}$, a time domain $\mathcal{T}$ and an input feature matrix $\mathcal{X}$. It can be characterized by a sequence of time-stamped edges $\mathcal{G} = \{(v_i, v_j, \tau)_n:n=1,2, \ldots,\left | \mathcal{E}\right | \}$. Here, each tuple $(v_i, v_j, \tau)$ denotes a distinct interaction between nodes $v_i$ and $v_j$ at time $\tau \in \mathcal{T}$, with $\left | \mathcal{E}\right |$ representing the total number of interactions in the dynamic graph. Given the dynamic graph $\mathcal{G}$, we learn a model with parameter $\theta$ to capture the temporal evolution of the graph. The temporal representations encoded by the learned model $\theta$ can be used for different tasks such as node classification, link prediction and graph classification.

\begin{figure*}[ht]
	\centering
	\includegraphics[scale=0.64]{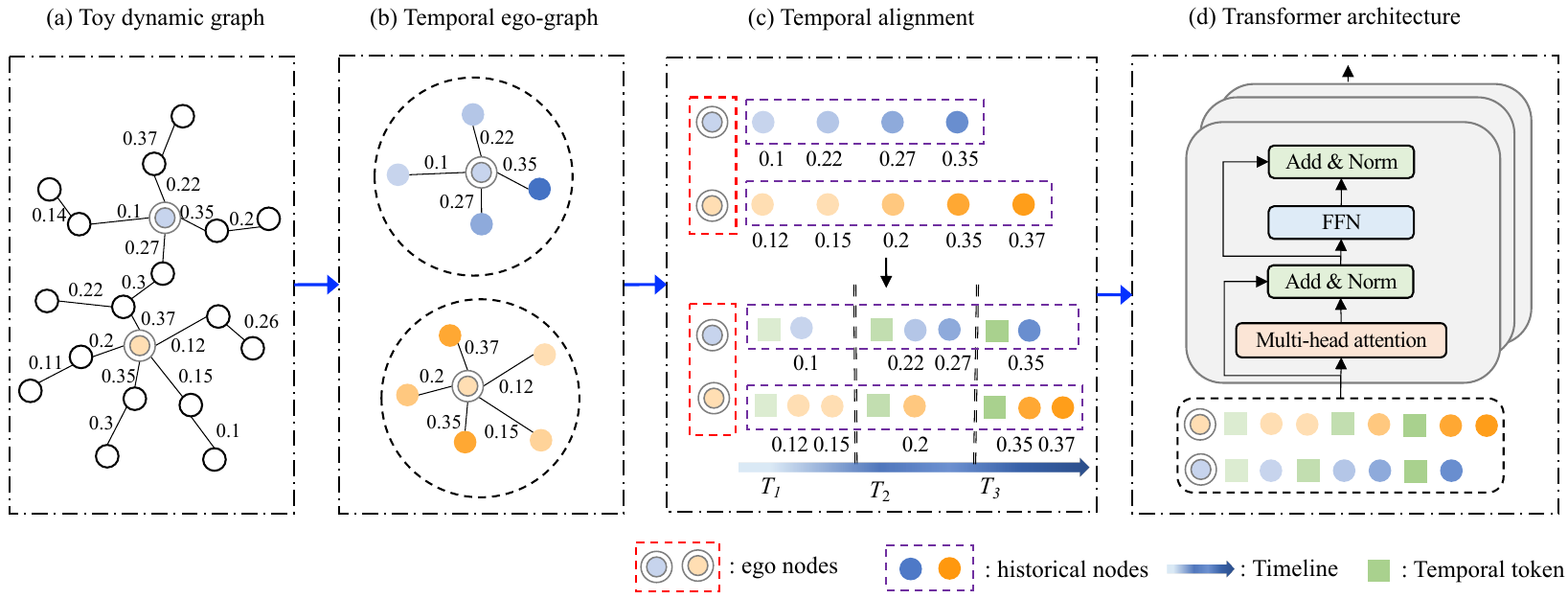}
	\vspace{-0.6cm}
	\caption{Overall framework of SimpleDyG. Best viewed in color. The numerical values adjacent to the links in (a) and (b), as well as beneath the nodes in (c), represent the time elapsed from the beginning, indicating the moments at which the links emerge (ranging from 0 to 1). The color intensity of nodes in the historical sequence represents the time span, where darker colors signify a longer duration, while lighter colors indicate a shorter duration.}
\label{framework}
\end{figure*}

\subsection{Transformer Architecture}

The standard Transformer architecture comprises two main components: the multi-head self-attention layers (MHA) and the position-wise feed-forward network (FFN). In the following part, we will briefly introduce these blocks.

We represent an input sequence as $\textbf{H}=\langle  \textbf{h}_1, \dots, \textbf{h}_N \rangle  \in \mathbb{R}^{N \times d} $, where $\textbf{h}_i$ is the hidden representation for token $i$, and $d$ is the dimension of the representations. The MHA module projects \textbf{H} to a triplet $(\textbf{Q},\textbf{K}, \textbf{V})$, as follows.
\begin{equation}
   \textbf{Q}=\textbf{HW}_Q, \textbf{K}=\textbf{HW}_K, \textbf{V}=\textbf{HW}_V  ,
\end{equation}
where ${\textbf{W}_Q} \in  \mathbb{R}^{d \times d_K}$, ${\textbf{W}_K} \in  \mathbb{R}^{d \times d_K}$, ${\textbf{W}_V} \in  \mathbb{R}^{d \times d_V}$ are learnable weights, with $d_K=d_V=d/H$.   
Overall, $H$ such projections are performed, resulting in $(\textbf{Q}_h, \textbf{K}_h, \textbf{V}_h)$ for $1\le h\le H$. The self-attention operation is then applied 
to each triplet:
\begin{align}
\text{head}_h&=\textsc{Softmax}\left(\textstyle\frac{\textbf{Q}_h\textbf{K}_h^T}{\sqrt{d_{K}}}\right)\textbf{V}_h, \\
    \textsc{MHA}(\textbf{H}) &= \textsc{Concat}(\text{head}_1,\ldots, \text{head}_H)\textbf{W}_O,
 \end{align}
where ${\textbf{W}_O} \in  \mathbb{R}^{d \times d}$ is a learnable weight matrix.

The output of the MHA module is then passed through a feed-forward network layer followed by residual connection \cite{he2016deep} and layer normalization (LN) \cite{ba2016layer}. Finally, the output of the $\textit{l}^\text{th} $ layer $\textbf{H}^l$ is computed as follows:
\begin{align}
\widehat{\textbf{H}}^{l} &=\textsc{LN}(\textbf{H}^{l-1}+\textsc{MHA}(\textbf{H}^{l-1})),\\
\textbf{H}^l &=\textsc{LN}(\widehat{\textbf{H}}^{l}+ \textsc{FFN}(\widehat{\textbf{H}}^{l})) .
\end{align}

%% file: method.tex
\section{Proposed Approach}

The overall framework of SimpleDyG is illustrated in Figure~\ref{framework}. Our framework is applied to a dynamic graph $\mathcal{G}$ (Figure~\ref{framework}(a)), where multiple temporal links emerge at various time points. In order to capture the dynamic evolution, we begin by extracting a \textit{temporal ego-graph} for each ego-node, which contains the entire historical interactions as shown in Figure~\ref{framework}(b). These temporal graphs are subsequently transformed into sequences while preserving their chronological order. To incorporate temporal alignment among different ego-graphs, we segment the timeline into various time spans with the same temporal interval as in  Figure~\ref{framework}(c). Then, we add \textit{temporal tokens} into the ego-sequence to identify different time spans. Finally, these sequences are fed into a Transformer architecture to facilitate various downstream tasks.

\subsection{Temporal Ego-graph}

The strategy of mapping a dynamic graph into a sequence of tokens is crucial for utilizing the Transformer architecture for dynamic graph modeling.  In this work, we regard nodes in the dynamic graphs as input tokens, which is a common approach in Transformer models for graphs. Besides, to preserve the historical interactions of all the nodes while ensuring scalability, we extract a temporal ego-graph for each node in the dynamic graph. Each ego-graph serves as the receptive field of its ego-node, which is mapped into a sequence to capture the structural and temporal evolution of the ego-node.

Specifically, we denote $v_i \in \mathcal{V}$ as an ego-node in the dynamic graph $\mathcal{G}$. We extract the historically interacted nodes for $v_i$ and construct an temporal ego-graph centered at $v_i$.
Formally, we denote the \textit{temporal ego-graph} for the ego-node $v_i$ as a chronologically ordered sequence $w_i=\langle v_i^1, v_i^2 \dots v_i^{|w_i|} \rangle$, where $|w_i|$ is the length of the sequence.
Note that $\forall 1\le k< k'\le |w_i|$, $v_i^k$ and $v_i^{k'}$ represent some historical interactions $(v_i,v_i^k,\tau)$ and $(v_i,v_i^{k'},\tau')$, respectively, such that $\tau \le \tau'$.

To better model the patterns within the sequences, we follow similar practices as in natural language processing and include some special tokens designed for our task. During training, the input sequence $x_i$ and the ground-truth output sequence $y_i$ are constructed as follows\footnote{Special tokens in the beginning and at the end such as ``$\langle \text{|\textit{endoftext}|} \rangle$'' are omitted for easy illustration.}.
\begin{equation}
\begin{aligned}
x_i &= \langle \text{|\textit{hist}|} \rangle, v_i, v_i^1,\dots v_i^{|w_i|}, \langle \text{|\textit{endofhist}|}\rangle ,\\
y_i &= \langle \text{|\textit{pred}|}\rangle, v_i^{{|w_{i}|+1}}, \dots, v_i^{{|w_{i}|}+z} \langle \text{|\textit{endofpred}|}\rangle,
\end{aligned}
\end{equation}
where the ``$\langle \text{|\textit{hist}|} \rangle$'' and ``$\langle \text{|\textit{endofhist}|} \rangle$'' are special tokens indicating the start and end of the input historical sequence, and ``$\langle \text{|\textit{pred}|} \rangle$'' and ``$\langle \text{|\textit{endofpred}|} \rangle$'' are reserved for signalling the prediction of the next nodes following the input sequence. Specifically, the model will halt its predictions once the end special token is generated, enabling automatic decisions on the number of predictions at a future time point. Potentially, by devising appropriate special tokens, our framework can also support other operations such as link deletion, as illustrated in Appendix~\ref{deletion}.

\subsection{Temporal Alignment}
In the original Transformer, the input sequence is treated as a sequence of tokens, and the model captures the relationships between these tokens based on their relative positions in the sequence. For dynamic graph modeling, the positions in the sequence represent the temporal order. However, it inherently lacks the capability to account for a universal time domain across sequences, to synchronize the time interval and frequency information.

We introduce a straightforward yet effective strategy to incorporate temporal alignment into the input sequences of the Transformer. 
First, we segment the time domain $\mathcal{T}$ into coarse-grained time steps, where the intervals between two consecutive time steps are equal, such as one week or one month, determined by dataset characteristics. It is important to note that our approach differs from discrete-time approaches: Within each time step, we consider the precise temporal order of each event. In contrast, discrete-time approaches treat each time step as a static snapshot. Next, we incorporate special \textit{temporal tokens} into the sequences, which explicitly denote different time steps that are globally recognized across all sequences. Suppose we split the time domain $\mathcal{T}$ into $T$ time steps, where the input sequence includes the first $T-1$ time steps and the output sequence covers the last time step. Then, the sequences of the ego-node $i$ can be  denoted as follows:
\begin{align}
 x_i' &= \langle \text{|\textit{hist}|}\rangle, v_i, \langle \text{|\textit{time1}|}\rangle,  S_i^1, \dots \langle \text{|\textit{timeT-1}|}\rangle, S_i^{T-1}, \langle \text{|\textit{endofhist}|}\rangle, \\   
y_i' &= \langle \text{|\textit{pred}|}\rangle, \langle \text{|\textit{timeT}|}\rangle, S_i^{T} \langle \text{|\textit{endofpred}|}\rangle, \\
S_i^t &= \langle v_i^{t,1}, v_i^{t,2} \dots v_i^{t,|S_i^t|} \rangle.
\end{align}
Here $S_i^t$ represents the historical sequence of node $v_i$ at time step $t$ with length $|S_i^t|$. In particular, ``$\langle \text{|\textit{time1}|} \rangle$'', $\dots$, ``$\langle \text{|\textit{timeT}|} \rangle$'' are temporal tokens that serve as indicators of temporal alignment, enabling the model to recognize and capture temporal patterns in different sequences. The temporal tokens enhances the Transformer's ability to understand the dynamics across the entire graph.


\subsection{Training objective}

A training instance is formed by concatenating the input $x$ and output $y$ as $[x;y]$. We denote it as $R = \langle r_1, r_2, \cdots, r_{\left | R \right |} \rangle $ with $\left | R \right |$ tokens. For a given training instance in this format, we follow the typical  masking strategy: During the prediction of the $i$-th token, only the input sequence up to position $i-1$, denoted by $R_{<i}$, is taken into account, while the subsequent tokens are subject to masking. The joint probability of the next token is calculated as follows:
\begin{align}
p(R) = \prod_{i=1}^{\left | R \right |}p(r_i|R_{<i}),
\label{equa-joint-prob}
\end{align}
where $p(r_i|R_{<i})$ is the probability distribution of the token to be predicted at step $i$ conditioned on the tokens $R_{<i}$. It is computed as
\begin{align}
    p(r_i|R_{<i}) = \textsc{LN}({\textbf{R}^l_{<i}})\textbf{W}_\text{vocab},
\label{equa-prob}
\end{align}
where \textsc{LN} means layer normalization, ${\textbf{R}^L_{<i}}$ denotes the hidden representation of the historically generated tokens before step $i$, which is obtained from the last layer of the Transformer, and $\textbf{W}_\text{vocab}$ is a learnable matrix aiming to compute the probability distribution across the vocabulary of nodes in the graph.

Given a set of training instances $\mathcal{R}$, the loss function for training the model with parameters $\theta$ is defined as the negative log-likelihood over $\mathcal{R}$, as follows:
\begin{align}
\mathcal{L} = -\sum_{R\in\mathcal{R}}\sum_{i=1}^{|R|}\log p_{\theta}(r_{i}|R_{<i}).
\label{equa}
\end{align}

We outline the training procedure of SimpleDyG in Algorithm~\ref{alg}. For each prediction step $i$ of one training instance, the hidden representations of the generated sequence   $\textbf{R}_{<i}$ are used for predicting the next token. The joint probability of the next token is computed using Equations~\ref{equa-joint-prob} and \ref{equa-prob}. Our model is trained using the Adam optimizer with a loss function based on the negative log-likelihood in Equation~\ref{equa}. 

\begin{algorithm}[t]
	
	\KwInput{Dynamic graph $\mathcal{G}=(\mathcal{V},\mathcal{E}, \mathcal{T}, \mathcal{X})$, \\
 \quad \quad \quad training instances $\mathcal{R}$
             } 
	\KwOutput{Model with parameters $\theta$}
	
	initialize $\theta$ 
	
	\While{not converged}
	{ 
		sample a batch of instances $\mathcal{B}$  from $\mathcal{R}$

        \For {each instance $R = \langle r_1, r_2, \cdots, r_{\left | R \right |} \rangle \in \mathcal{B}$}
            {
                \While{step $i < {\left | R \right |} $}
                {

                \tcc{prediction steps for one instance}
                
                    calculate the representation $\textbf{R}_{<i}$ for $R_{<i}$
            
                    calculate the joint probability by Eqs.~\ref{equa-joint-prob} and \ref{equa-prob}

                    calculate the loss by Eq.~\ref{equa}
                            
                }
            }
        update $\theta$ via backpropagation
	}
    \Return $\theta$ 
	\caption{Training Procedure of SimpleDyG}
	\label{alg}
\end{algorithm}

%% file: exp.tex
\section{Experiments}
In this section, we conduct extensive experiments on four public datasets across different domains, with comparison to the state-of-the-art baselines and detailed analysis of the model performance.

\subsection{Experimental Setup}
\vpara{Datasets.} To evaluate the performance of our proposed method, we conducted experiments on four datasets from various domains, including the communication network UCI \cite{panzarasa2009patterns}, the rating network ML-10M \cite{harper2015movielens}, the citation network Hepth \cite{leskovec2005graphs}, and the multi-turn conversation dataset MMConv \cite{liao2021mmconv}. The detailed statistics of all datasets after preprocessing are presented in Table \ref{tab:dataset}.

\begin{table}[t]
    \centering
    \small
    \caption{Dataset statistics}
    \label{tab:dataset}
    \begin{tabular}{cccccl}
    \toprule
    Dataset &  UCI &ML-10M &Hepth & MMConv  \\
    \midrule
    Domain & Social  &Rating   &Citation &Conversation  \\
    {\#} Nodes &1,781  &15,841   & 4,737  &7,415   \\
    {\#} Edges &16,743   &48,561    &14,831  &91,986      \\ 
  \bottomrule
\end{tabular}
\end{table}

\textbf{UCI} \cite{panzarasa2009patterns}: It represents a social network in which edges represent messages exchanged among users. For temporal alignment, we employ 13 time steps following previous work \cite{sankar2020dysat}.

\textbf{ML-10M} \cite{harper2015movielens}: We utilize the ML-10M dataset from MovieLens comprising user-tag interactions, where the edges connect users to the tags they have assigned to specific movies. For temporal alignment, we employ 13 time steps following previous work \cite{sankar2020dysat}.

\textbf{Hepth} \cite{leskovec2005graphs}: It is a citation network for high-energy physics theory. For temporal alignment, we extract 24 months of data from this dataset and split them into 12 time steps based on the wall-clock timestamps. Note that this dataset contains new emerging nodes as time goes on. We use the word2vec model \cite{mikolov2013distributed}  to extract the input feature for each paper based on the abstract. 

\textbf{MMConv} \cite{liao2021mmconv}: It contains a multi-turn task-oriented dialogue system that assists users in discovering places of interest across five domains. Leveraging this rich annotation, we represent the dialogue as a dynamic graph, a widely adopted strategy in task-oriented dialogue systems. For temporal alignment, we employ 16 time steps, each corresponding to a distinct turn in the conversation.

\begin{table*}[ht]
	\centering
	\addtolength{\tabcolsep}{-1pt}

    \caption{\label{result} Performance of dynamic link prediction by SimpleDyG and the baselines on four datasets. (In each column, the best result is bolded and the runner-up is \underline{underlined}.)
    }
        \vspace{-0.1cm}
 
    	\begin{tabular}{@{}c|cc|cc|cc|cc@{}}
		
		\toprule
		& \mc{2}{c|}{UCI} & \mc{2}{c|}{ML-10M} & \mc{2}{c|}{Hepth} & \mc{2}{c}{MMConv} \\
		\cmidrule{2-9}
		& {$NDCG@5$}     & {$Jaccard$}   &{$NDCG@5$}  & {$Jaccard$}   &{$NDCG@5$} & {$Jaccard$}     &{$NDCG@5$} & {$Jaccard$}    \\		
		
		\midrule		
  
		 DySAT \cite{sankar2020dysat} & 0.010$\pm$0.003    & 0.010$\pm$0.001     & 0.058$\pm$0.073   &0.050$\pm$0.068      &  0.007$\pm$0.002   &  0.005$\pm$0.001 &0.102$\pm$0.085    &\underline{0.095$\pm$0.080}      \\
		
		 EvolveGCN \cite{pareja2020evolvegcn}  &0.064$\pm$0.045     & 0.032$\pm$0.026     & \underline{0.097$\pm$0.071}   &\underline{0.092$\pm$0.067}       & 0.009$\pm$0.004    & 0.007$\pm$0.002 &0.051$\pm$0.021    &0.032$\pm$0.017     \\
        \midrule
         DyRep \cite{trivedi2019dyrep}  & 0.011$\pm$0.018    & 0.010$\pm$0.005     & 0.064$\pm$0.036  &0.038$\pm$0.001     &0.031$\pm$0.024     & 0.010$\pm$0.006  &0.140$\pm$0.057    & 0.067$\pm$0.025      \\

         JODIE \cite{kumar2019predicting}  & 0.022±0.023    & 0.012±0.009    & 0.059±0.016  & 0.020±0.004    & 0.031±0.021 & 0.011±0.008 & 0.041±0.016    & 0.032±0.022 \\ 
        
		 TGAT \cite{xu2020inductive}  & 0.061$\pm$0.007    & 0.020$\pm$0.002     &0.066$\pm$0.035   &0.021$\pm$0.007  & \underline{0.034$\pm$0.023}    & \underline{0.011$\pm$0.006}  &0.089$\pm$0.033    & 0.058$\pm$0.021    \\

         TGN \cite{rossi2020temporal} &0.041±0.017  &0.011±0.003    & 0.071±0.029  & 0.023±0.001    & 0.030±0.012 & 0.008±0.001   & 0.096 ±0.068   & 0.066±0.038  \\

         TREND \cite{wen2022trend}  &0.067±0.010	&0.039±0.020	&0.079±0.028	&0.024±0.003	&0.031±0.003	&0.010±0.002	&0.116±0.020	&0.060±0.018  \\

         GraphMixer \cite{cong2022we}  & \textbf{0.104$\pm$0.013}     & \underline{0.042$\pm$0.005 }    &0.081$\pm$0.033   &0.043$\pm$0.022       &0.011$\pm$0.008  & 0.010$\pm$0.003   &\underline{0.172$\pm$0.029}    &0.085$\pm$0.016    \\
		\midrule
	       
		 SimpleDyG  & \textbf{0.104$\pm$0.010}    & \textbf{0.092$\pm$0.014}     & \textbf{0.138$\pm$0.009}   & \textbf{0.131$\pm$0.008}      & \textbf{0.035$\pm$0.014}    & \textbf{0.013$\pm$0.006}    & \textbf{0.184$\pm$0.012}   & \textbf{0.169$\pm$0.010}    \\ 
  
		\bottomrule
	\end{tabular}
	
\end{table*}

\vpara{Baselines.} We compare SimpleDyG with baselines in two categories: (1) discrete-time approaches: DySAT \cite{sankar2020dysat} and EvolveGCN \cite{pareja2020evolvegcn}; (2) continuous-time approaches: DyRep \cite{trivedi2019dyrep}, JODIE \cite{kumar2019predicting}, TGAT \cite{xu2020inductive}, TGN \cite{rossi2020temporal}, TREND \cite{ wen2022trend} and GraphMixer \cite{cong2022we}.

\begin{itemize}[leftmargin=*]
	\item \textbf{DySAT} \cite{sankar2020dysat} leverages joint structural and temporal self-attention to learn the node representations at each time step. 

    \item \textbf{EvolveGCN} \cite{pareja2020evolvegcn} employs RNNs to evolve graph convolutional network parameters. 

\item \textbf{DyRep} \cite{trivedi2019dyrep} utilizes a two-time scale deep temporal point process model to capture temporal graph topology and node activities. 

\item \textbf{JODIE} \cite{kumar2019predicting} focuses on modeling the binary interaction among users and items by two coupled RNNs. A projection operator is designed to predict the future node representations at any time. 

\item \textbf{TGAT} \cite{xu2020inductive}  employs temporal graph attention layers and time encoding techniques to aggregate temporal-topological features. 

\item \textbf{TGN} \cite{rossi2020temporal} combines the memory modules and message passing to maintain the dynamic representations. This model also adopts time encoding and temporal graph attention layers.

\item \textbf{TREND} \cite{wen2022trend} exploits Hawkes process-based GNNs for dynamic graph modeling. It integrates event and node dynamics to capture the individual and collective characteristics of events.

\item \textbf{GraphMixer} \cite{cong2022we} relies on MLP layers and neighbor mean-pooling to learn the edge and node encoders. An offline time encoding function is adopted to capture the temporal information.
\end{itemize}

\vpara{Implementation Details.} We evaluate the performance of SimpleDyG on the link prediction task. Given the ego-nodes, the objective of the link prediction task is to predict the possible linked nodes at time step $T$. For all the datasets, we follow the previous setting \cite{cong2022we} by treating the dynamic graphs as undirected graphs. We split each dataset into training/validation/testing based on the predefined time steps. We choose the data at the last time step ${T}$ as the testing set, while the data at time step ${T-1}$ serves as the validation set, with the remaining data for training.  We tune the hyperparameters for all methods on the validation set. All experiments are repeated ten times, and we report the averaged results with standard deviation.
We provide further implementation details and hyperparameter settings in Appendices~\ref{imp} and \ref{hyper}.


\vpara{Evaluation Metrics.} In our evaluation, we carefully select metrics catered to our specific task. The goal of the link prediction task is to 
predict a set of nodes linked to each ego-node at a given time step. Notably, our SimpleDyG model predicts a node sequence, with each prediction influenced by the prior ones until the generation of an end token. In contrast, the baseline models make simultaneous predictions of entire ranking sequences for each ego-node. To evaluate the ranking performance and the similarity between predicted and ground-truth node sets, we employ two key metrics: \textit{NDCG@5} and \textit{Jaccard} similarity. \textit{NDCG@5} is a well-established metric commonly used in information retrieval and ranking tasks \cite{wang2019neural}, aligning with our objective of ranking nodes and predicting the top nodes linked to an ego-node. On the other hand, \textit{Jaccard} similarity is valuable for quantifying the degree of overlap between two sets \cite{jaccard1901etude}, measuring the similarity between predicted nodes and the ground-truth nodes associated with the ego-node. Specifically, for the baseline models, we choose the top \textit{k} nodes ($k=1,5,10,20$) as the predicted set, as they are not generative models and cannot determine the end of the prediction. We then select the maximum Jaccard similarity value across different \textit{k}'s as the final Jaccard similarity score. This evaluation strategy ensures a thorough and fair assessment of the baselines in comparison to SimpleDyG.

\begin{table*}[ht]
	\centering
    \caption{\label{special} Impact of special tokens in SimpleDyG across four datasets.}
       \vspace{-0.1cm}
    	\begin{tabular}{@{}c|cc|cc|cc|cc@{}}
		
		\toprule
		& \mc{2}{c|}{UCI} & \mc{2}{c|}{ML-10M} & \mc{2}{c|}{Hepth} & \mc{2}{c}{MMConv} \\
		\cmidrule{2-9}
		& {$NDCG@5$}     & {$Jaccard$}   &{$NDCG@5$}  & {$Jaccard$}   &{$NDCG@5$} & {$Jaccard$}     &{$NDCG@5$} & {$Jaccard$}    \\		
		
		\midrule		
		
		 \textit{SimpleDyG}  & \underline{0.104$\pm$0.010 }   & \underline{0.092$\pm$0.014 }   & \textbf{0.138$\pm$0.009}   & \textbf{0.131$\pm$0.008}    & \underline{0.035$\pm$0.014}    & \underline{0.013$\pm$0.006}  & \textbf{0.184$\pm$0.012}   & \underline{0.169$\pm$0.010}       \\

		 \textit{same special}  & \textbf{0.113$\pm$0.007}    & \textbf{0.095$\pm$0.010}     & \underline{0.085$\pm$0.046}   & \underline{0.079$\pm$0.043}    & 0.027$\pm$0.014   & 0.009$\pm$0.005  & \underline{0.179$\pm$0.013}  & \textbf{0.170$\pm$0.010}     \\   
   
         \textit{no special}  & 0.041$\pm$0.025    & 0.020$\pm$0.011    & 0.006$\pm$0.009   & 0.006$\pm$0.009     & \textbf{0.096$\pm$0.016	}    & \textbf{0.025$\pm$0.006}  & 0.01$\pm$0.008   & 0.008$\pm$0.007      \\    
		\bottomrule
	\end{tabular}
\end{table*}

\subsection{Performance Comparison to Baselines}
We report the results of all methods under \textit{NDCG@5} and \textit{Jaccard} across four diverse datasets in Table~\ref{result}. Generally speaking, our method outperforms all the baselines on all datasets, and we make the following key observations.

First, we find that continuous-time approaches generally perform better than discrete ones across a wide range of scenarios, indicating the important role of time-related information in dynamic graph analysis. Notably, continuous-time baselines such as GraphMixer exhibit superior performance. This superiority can be mainly attributed to the simple MLP-Mixer architecture, which makes it easier to capture long-term historical sequences with lower complexity. In contrast, other models like DyRep, TGAT, and TGN, which rely on complex designs such as GNNs and GATs, display subpar performance. This phenomenon stems from the inherent limitations of GNNs and GATs in capturing distant relationships or broader historical contexts within predefined time windows.

Second, in inductive scenarios like the Hepth dataset, where the ego-nodes in the testing data are newly emerged nodes, continuous-time models that employ a  GNN-based backbone exhibit superior performance compared to GraphMixer. To be able to handle new nodes, the initial node features are constructed using word2vec, which might be relatively coarse. Since GraphMixer predominantly relies on an MLP-based architecture, it may encounter challenges given the coarse-grained initial features. Conversely, GNN-based methods integrate structural information with these features, thereby empowering them to excel in the inductive scenario. Nevertheless, in our Transformer-based model, there is the added advantage of modeling long-range dependencies, resulting in consistently better performance of SimpleDyG. 


\subsection{Effect of Extra Tokens}
\begin{table*}[t]
	\centering
	\renewcommand*{\arraystretch}{1}
    \caption{\label{align} Impact of different temporal alignment designs on the four datasets.}
    \vspace{-0.1cm}
    	\begin{tabular}{@{}c|cc|cc|cc|cc@{}}
		
		\toprule
	&	 \mc{2}{c|}{UCI} & \mc{2}{c|}{ML-10M} & \mc{2}{c|}{Hepth} & \mc{2}{c}{MMConv} \\
		\cmidrule{2-9}
	&	 {$NDCG@5$}     & {$Jaccard$}   &{$NDCG@5$}  & {$Jaccard$}   &{$NDCG@5$} & {$Jaccard$}     &{$NDCG@5$} & {$Jaccard$}    \\		
		
		\midrule		
		
		 \textit{SimpleDyG}  & \underline{0.104$\pm$0.010}   & \textbf{0.092$\pm$0.014}    & 0.138$\pm$0.009  & 0.131$\pm$0.008      & 0.035$\pm$0.014    & 0.013$\pm$0.006  & 0.184$\pm$0.012  & 0.169$\pm$0.010     \\ 

		 \textit{same time}  & 0.090$\pm$0.013	& 0.083$\pm$0.012	& \textbf{0.147$\pm$0.001}	& \textbf{0.139$\pm$0.001}		& \textbf{0.046$\pm$0.009}	& \underline{0.017$\pm$0.004} & \underline{0.240$\pm$0.031}	& \underline{0.212$\pm$0.025} \\

         \textit{no time} & \textbf{0.111$\pm$0.015}	& \underline{0.091$\pm$0.014} &0.117$\pm$0.062 & 0.111$\pm$0.059				& \underline{0.045$\pm$0.007}	& \textbf{0.018$\pm$0.003}   & \textbf{0.260$\pm$0.019}	& \textbf{0.237$\pm$0.016}    \\    
		\bottomrule
	\end{tabular}
\end{table*}

We design extra tokens to make the vanilla Transformer architecture more suitable for dynamic graph modeling. To assess their effectiveness, we conduct an in-depth analysis of these token designs, including \textit{special tokens} that signal the input and output, and the \textit{temporal tokens} that align different sequences.

\vpara{Impact of \textit{special tokens}.} The \textit{special tokens} include the start and end of the historical sequence (``$\langle \text{|\textit{hist}|} \rangle$'' and ``$\langle \text{|\textit{endofhist}|} \rangle$''), as well as the predicted sequence (``$\langle \text{|\textit{pred}|} \rangle$''and ``$\langle \text{|\textit{endofpred}|} \rangle$'').  To comprehensively evaluate their effect across diverse scenarios, we examine two variants of SimpleDyG: (1) \textit{same special}, where we use the same \textit{special tokens} for input and output. (2) \textit{no special}, where we entirely remove all special tokens from each instance. We show the results in Table~\ref{special} and make the following observations. 

In general, \textit{special tokens} enhance the link prediction performance across different datasets. Furthermore, the differences between the \textit{same special} and original SimpleDyG tend to be small. 
However, an interesting finding emerges in the case of the Hepth dataset, where the \textit{no special} variant yields the best performance. It can be explained by the specific characteristic of the citation graph. In the testing data of Hepth, the ego-nodes are newly emerged nodes, representing the newly published papers. Consequently, the input samples lack any historical information, leaving the distinction between the history and the predicted irrelevant. 

\vpara{Impact of \textit{temporal tokens}.} 
To evaluate the impact of \textit{temporal tokens}, we compare the performance of SimpleDyG with two ablated variants: (1) \textit{same time}, where we do not distinguish specific time steps and employ the same \textit{temporal tokens} for each time step, and (2) \textit{no time}, in which we entirely remove all temporal tokens from all sequences.

\begin{figure*}[ht]  
\subfigure[UCI]{
\centering
    \includegraphics[scale=0.73]{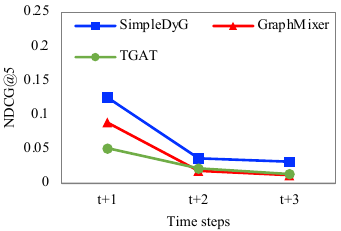}
    \label{uci}
}
\subfigure[ML-10M]{
    \centering
    \includegraphics[scale=0.73]{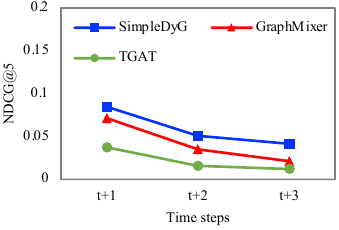}
    \label{ml-10m}
}
\subfigure[Hepth]{
    \centering
    \includegraphics[scale=0.73]{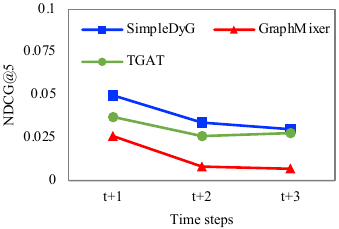}
    \label{hepth}
}
\subfigure[MMConv]{
    \centering
    \includegraphics[scale=0.73]{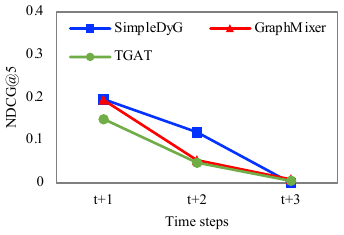}
    \label{mmconv}
}
\vspace{-0.3cm}
 \caption{Performance trends of multi-step prediction.}
 \label{multi-step}
\end{figure*}

\begin{figure*}[ht] 
\centering
\subfigure[Number of layers]{
    \centering
    \includegraphics[scale=0.85]{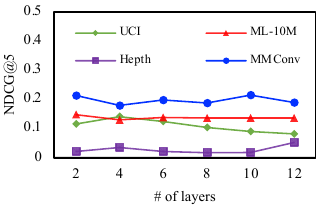}
    \label{layer}
    }
\subfigure[Number of heads]{
    \centering
    \includegraphics[scale=0.85]{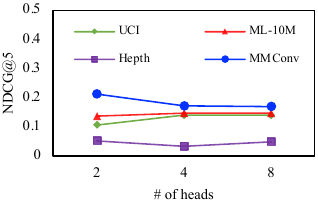}
    \label{head}
    }
\subfigure[Hidden dimension]{
    \centering
    \includegraphics[scale=0.85]{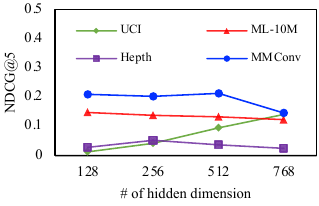}
    \label{hidden}
    }
    \vspace{-0.3cm}
\caption{Impact of hyperparameters.}
\label{para}
\end{figure*}

 The results are presented in Table~\ref{align}. It is surprising and interesting to observe performance improvement with less or no temporal alignment, particularly on the MMConv and Hepth datasets. We hypothesize that the citation relationship and the conversation among different ego-nodes do not strictly follow a universal temporal framework. Using the same temporal tokens (\emph{same time}) or none at all (\emph{no time}) allows the model to adapt more naturally to the temporal order. 
The temporal alignment plays a more important role for the UCI and ML-10M datasets. However, they show different trends with the \textit{same time} version. A potential reason is that, in UCI, the communication patterns between different users are sensitive to the segmentation into different time steps. Hence, \textit{same time} performs the worst, as it divides a sequence into time steps yet without time differentiation to align across the sequences, where the additional same tokens may confuse the model. On the other hand, \textit{no time} still retains the full temporal order, and thus perform better than \textit{same time}.

\subsection{Performance of Multi-step Prediction}

We evaluate the ability of SimpleDyG for multi-step prediction with the time steps ranging from $t$ to $t + \bigtriangleup t$, utilizing a model that has been trained on data up to time $t$. 
In our experiment, we set $\bigtriangleup t=3$. For SimpleDyG, we perform step-by-step prediction, conditioned on the results of previous steps. We compare to two competitive baselines, TGAT and GraphMixer. As both methods incorporate time information via time encoding, we employ the model trained at time $t$ to directly predict the links at time $t + \bigtriangleup t$. Their performance trends are plotted in Figure~\ref{multi-step}. 

We observe a natural decay in performance over time for all methods, as anticipated. However, SimpleDyG consistently outperforms the baselines as time progresses. This trend underscores the effectiveness of our Transformer architecture in modeling long-term dependencies in a dynamic graph. 

\subsection{Hyper-parameter Analysis}

We undertake an examination of the critical hyperparameter choices, taking into account the variations observed across different datasets. Specifically, we systematically explore the impact of several key hyperparameters, namely, the number of layers, the number of heads, and the hidden dimension size. These hyperparameters play a pivotal role in shaping the model's capacity and its ability to capture intricate patterns within dynamic graphs. We vary the value of each hyperparameters while keeping all other parameters constant. From Figure~\ref{para}, we draw some key observations as follows.

\begin{itemize}[leftmargin=*]
	\item \textbf{Number of layers}: The variance of performance under different numbers of layers is relatively small. This suggests that the choice of the number of layers in SimpleDyG has a more consistent impact across different datasets and scenarios. Generally speaking, two layers are typically sufficient for most cases. For inductive scenarios such as the Hepth dataset, it is advisable to use more layers to effectively capture the evolving graph structure. 

\item \textbf{Number of heads}: For the number of attention heads, we find that using either 2 or 4 heads is generally suitable for a wide range of scenarios. These settings provide a good balance between performance and computational efficiency.

\item \textbf{Hidden dimension size}: The choice of hidden dimension size depends on the complexity of the dataset. For datasets like movie ratings (e.g., ML-10M), a hidden dimension size of 128 is often adequate. However, for datasets involving more intricate interactions, such as communication networks or conversation datasets, it becomes necessary to use larger hidden dimension sizes like 256 or 512. Notably, the UCI dataset requires a hidden dimension of 768, which can be explained by the complexity and richness of the interactions among users within the dataset. 
\end{itemize}



%% file: app.tex
\section*{Appendices}

\section{Link Deletion Operation}
\label{deletion}

\begin{figure*}[ht]  
\subfigure[UCI]{
\centering
    \includegraphics[scale=0.4]{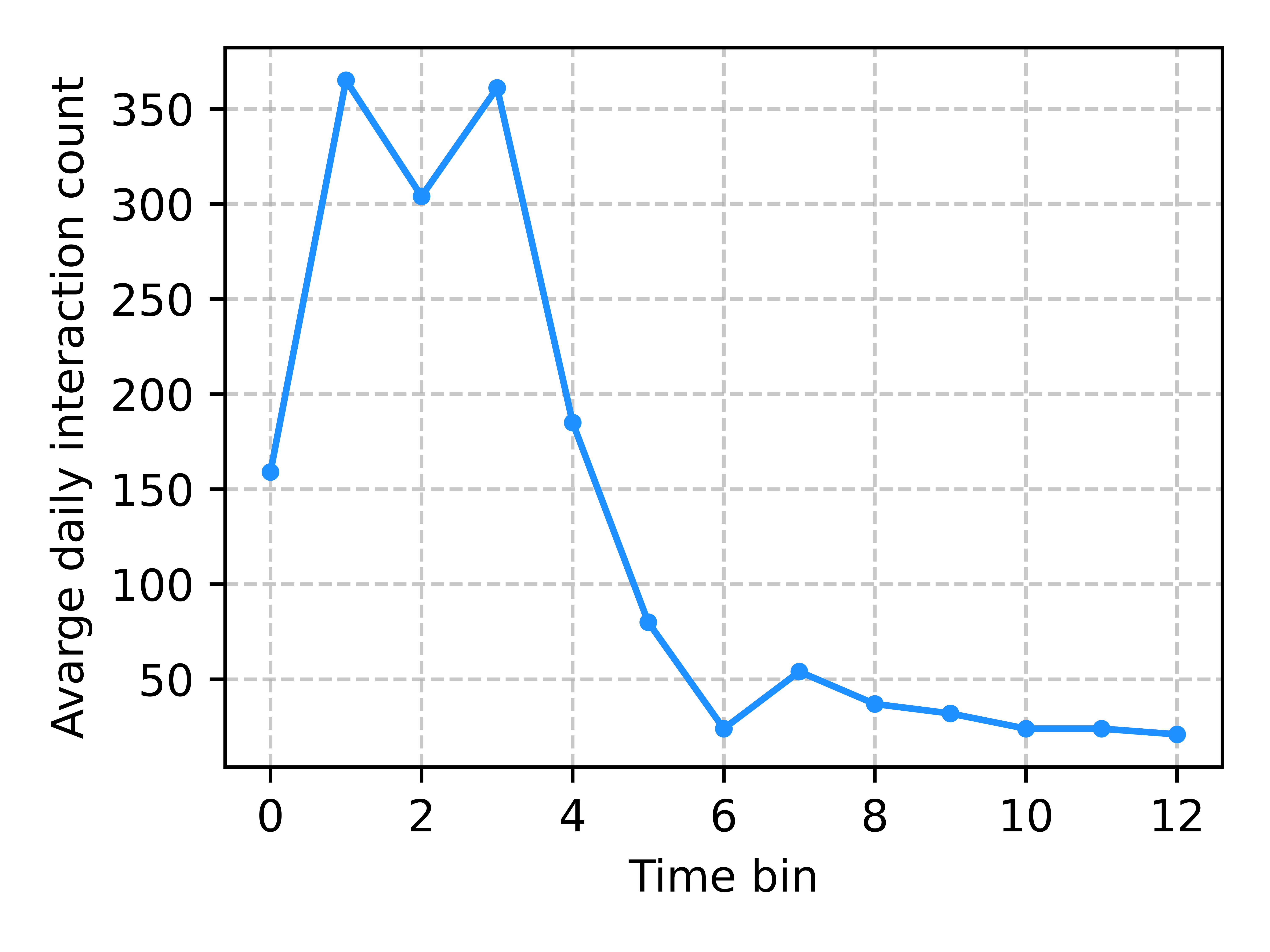}
    \label{uci}
}
\subfigure[ML-10M]{
    \centering
    \includegraphics[scale=0.4]{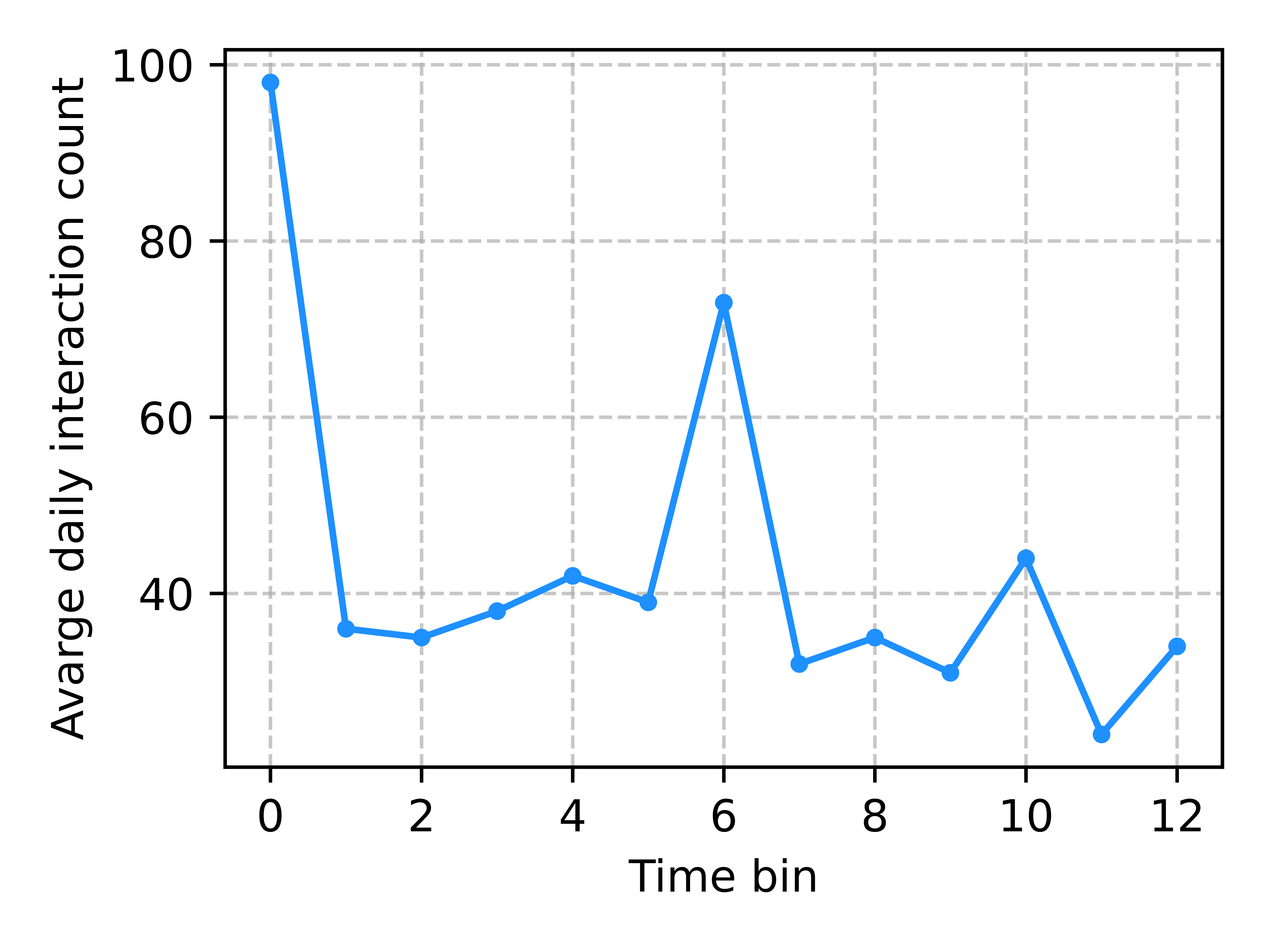}
    \label{ml-10m}
}
\subfigure[Hepth]{
    \centering
    \includegraphics[scale=0.4]{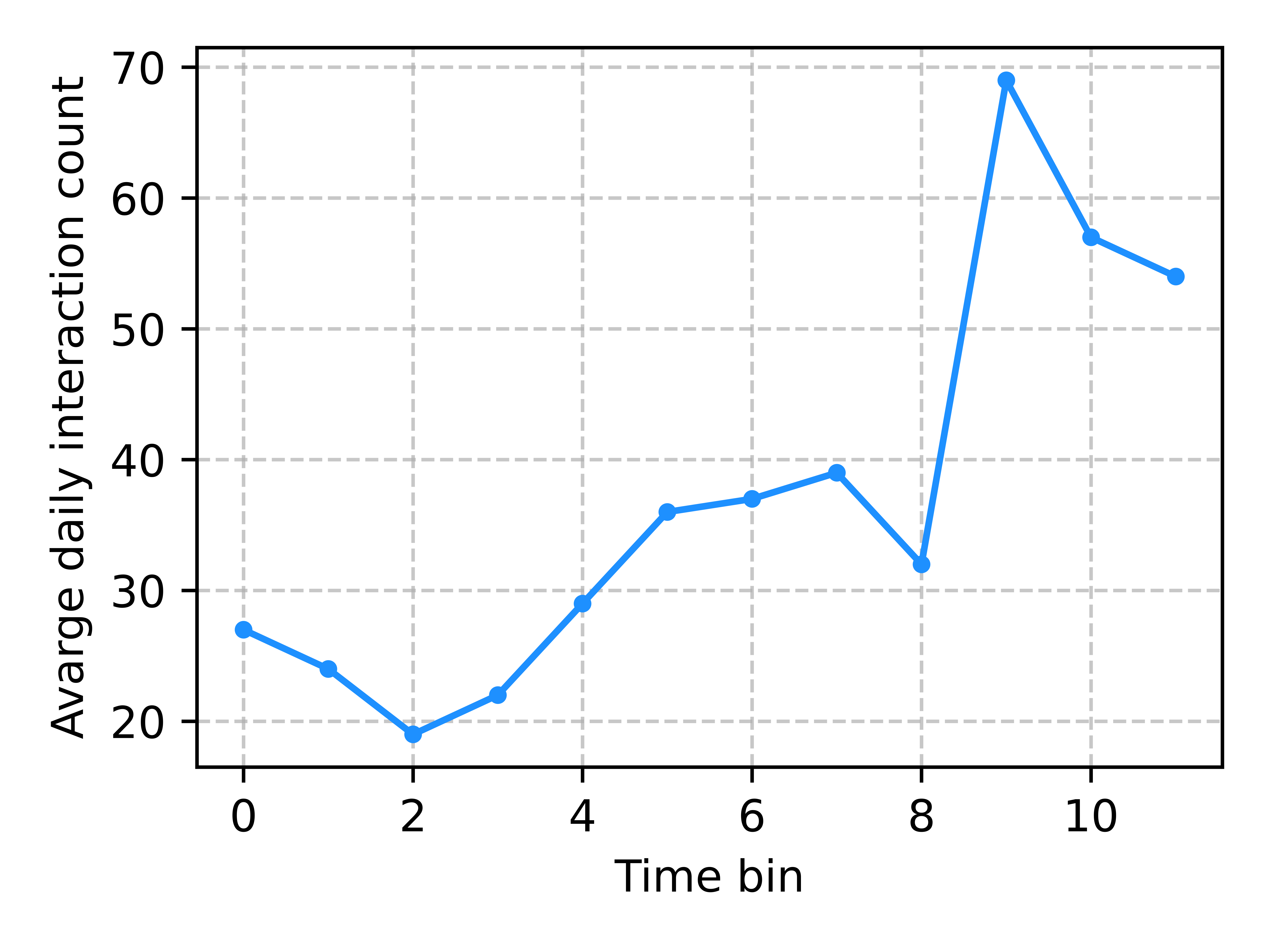}
    \label{hepth}
}
\subfigure[MMConv]{
    \centering
    \includegraphics[scale=0.4]{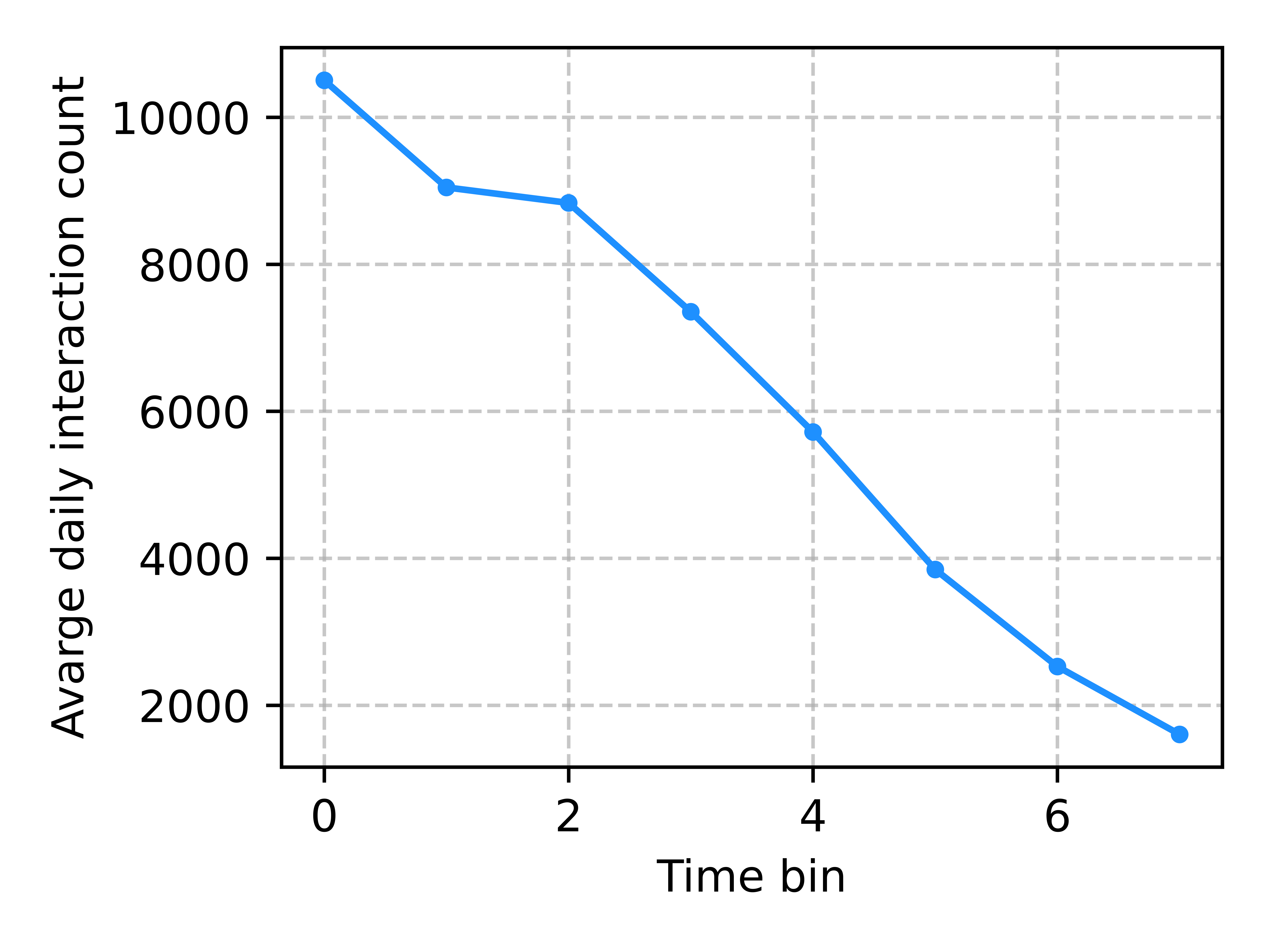}
    \label{mmconv}
}
\vspace{-0.3cm}
 \caption{The temporal pattern of each dataset.}
 \label{pattern}
 \vspace{-0.1cm}
\end{figure*}

\begin{table*}[htp]
    \centering
    \caption{Time Efficiency of Different Methods}
        \vspace{-0.2cm}
    \label{efficiency}
    \begin{tabular}{cccccccccc}
    \toprule
    Method &  DySAT &EvolveGCN &DyRep & JODIE & TGAT & TGN & TREND & GraphMixer &SimpleDyG  \\
    \midrule
    Time (s) & 12.24  & 11.89   & 6.4 & 6.25 & 18.54 & 8.05 & 7.45 & 6.89 & 6.21  \\
    
  \bottomrule
\end{tabular}
\end{table*}

Our framework can support the link deletion operation within the evolution of dynamic graphs. For example in the ML-10M dataset with links representing users’ ratings for the movies, assuming that users can delete the historical ratings. Suppose the historically rated movies of one user is ``$[a,b,c]$'', he/she may delete the rating of \textit{b}, then we can add a special token ``[del]'' to identify the deletion operation: ``$[a, [del], b, c]$''. It is worth noting that most works for dynamic graph modeling are associated with link addition with benchmark datasets mainly involving this type of operation.

\section{Additional Implement Details}
\label{imp}
Note that the implementation details of baseline approaches in their publicly released code are quite different. For instance, most of them regard the link prediction task as binary classification, where the objective is to determine the presence or absence of links between the positive pairs of nodes and randomly selected negative pairs. They either employ a binary cross-entropy loss to facilitate classifier learning or utilize logistic regression to train an additional classifier. To tailor these baselines to our specific task for a fair comparison, we adapt them into a ranking task and substitute the classifier loss with a pair-wise Bayesian personalized ranking (BPR) loss for all baselines.

\section{Hyperparameters Settings of Baselines}
\label{hyper}
Considering that we refine the loss of the baselines as BPR loss, we tune the important parameters of all baselines for all the datasets. For all baselines, we tune the parameter of hidden dimension with $\{16, 32, 64, 128, 256, 512\}$ for each dataset. For a fair comparison with our model, we don't set a historical window for discrete-time approaches and use all the historical data.

Some important parameters for each baseline are listed as follows:
For DySAT \cite{sankar2020dysat}, We set the self-attention layers and head to be 2 and 8, respectively. 
For EvolveGCN \cite{pareja2020evolvegcn}, the number of GCN layers is 1.
For DyRep \cite{trivedi2019dyrep}, the message aggregation layer is 2, and the number of neighbor nodes is 20.
For TGAT \cite{xu2020inductive} and TGN \cite{rossi2020temporal}, the number of graph attention heads is 2 and the attention layers are 1 and 2, respectively.
For GraphMixer \cite{cong2022we}, the number of MLP layers for UCI is 1 and 2 for other datasets. For a fair comparison, we set the historical length of each node to 1024, which is the same as our model.

\section{Statistical Temporal Pattern Analysis in Each Dataset}
We incorporate statistical analyses to understand the temporal patterns within each dataset. Specifically, we counted the number of interactions per day for the UCI, ML-10M, and Hepth datasets. For the conversation dataset (MMConv), we counted the number of interactions per turn. We split the dataset into several bins with each bin covering a range of time (10 days, 90 days, 60 days and two turns for UCI, ML-10M, Hepth and MMConv, respectively), and then calculate the average daily interaction counts in each bin. We show the results in Figure \ref{pattern}.


We observe sudden changes in interactions in the UCI and ML-10M datasets. While the Hepth and MMConv datasets generally show gradual change across the entire timeline. This indicates that in the UCI and ML-10M datasets, the interactions among different time slots show different patterns. Thus it is more important to conduct temporal alignment using temporal tokens which is consistent with our analysis for Table 4. For the Hepth and MMConv datasets, a simpler design for temporal alignment (e.g., same token or no token) is enough to achieve good performance.

\section{Time Complexity Analysis}
\label{time}
The time complexity of our SimpleDyG model is the same as the vanilla Transformer, which is $O(n^2)$
 where $n$ is the sequence length. We conduct time efficiency experiments about the training time per epoch of different methods on UCI dataset using a machine with a NVIDIA L40 GPU with 64 CPU cores. The results show that our method trains faster than or comparable to all baselines. The methods integrating temporal modeling (e.g., RNN, self-attention) with structural modeling (e.g., GNN, GAT) suffer from the complexity of these modules.